\documentclass[twocolumn,english,prl,aps,superscriptaddress]{revtex4-2}
\usepackage[T1]{fontenc}
\usepackage{color}
\usepackage{babel}
\usepackage{mathtools}
\usepackage{bm}
\usepackage{graphicx}
\usepackage{esint}
\usepackage{amsfonts}
\usepackage{graphicx}
\usepackage{float}
\usepackage{subfigure}
\usepackage{amsthm}
\usepackage{amssymb}
\usepackage[ruled, longend]{algorithm2e}

\usepackage[usenames,divipsnames,svgnames,table]{xcolor}
\usepackage[unicode=true,pdfusetitle,
bookmarks=true,bookmarksnumbered=false,bookmarksopen=false,
breaklinks=true,pdfborder={0 0 0},backref=false,colorlinks=true]{hyperref}
\hypersetup{linkcolor=NavyBlue,urlcolor=NavyBlue,citecolor=NavyBlue}
\usepackage{breakurl}

\begin{document}

\title{Advantage of quantum coherence in postselected metrology}
\author{Shao-Jie Xiong}
\affiliation{School of physics, Hangzhou Normal
University, Hangzhou 310036, China}
\affiliation{Zhejiang Institute of Modern Physics and Department of Physics,
Zhejiang University, Hangzhou, Zhejiang 310027, China}

\author{Peng-Fei Wei}
\affiliation{School of physics, Hangzhou Normal
University, Hangzhou 310036, China}

\author{Huang-Qiu-Chen Wang}
\affiliation{School of physics, Hangzhou Normal
University, Hangzhou 310036, China}

\author{Lei Shao}
\affiliation{Zhejiang Institute of Modern Physics and Department of Physics,
Zhejiang University, Hangzhou, Zhejiang 310027, China}

\author{Yong-Nan Sun}
\affiliation{School of physics, Hangzhou Normal
University, Hangzhou 310036, China}

\author{Jing Liu}
\email{liujingphys@hust.edu.cn}
\affiliation{National Precise Gravity Measurement Facility, MOE Key 
Laboratory of Fundamental Physical Quantities Measurement, School of Physics,
Huazhong University of Science and Technology, Wuhan 430074, China}

\author{Zhe Sun}
\email{sunzhe@hznu.edu.cn}
\affiliation{School of physics, Hangzhou Normal
University, Hangzhou 310036, China}

\author{Xiao-Guang Wang}
\email{xgwang1208@zju.edu.cn}
\affiliation{Zhejiang Institute of Modern Physics and Department of Physics,
Zhejiang University, Hangzhou, Zhejiang 310027, China}

\begin{abstract}
In conventional measurement, to reach the greatest accuracy of parameter estimation, all samples must be measured since 
each independent sample contains the same quantum Fisher information. In postselected metrology, postselection can concentrate 
the quantum Fisher information of the initial samples into a tiny post-selected sub-ensemble. It has been proven that this 
quantum advantage can not be realized in any classically commuting theory. In this work, we present that the advantage of 
postselection in weak value amplification (WVA) can not be achieved without quantum coherence. The quantum coherence of the 
initial system is closely related to the preparation costs and measurement costs in parameter estimation. With the increase 
of initial quantum coherence, the joint values of preparation costs and measurement costs can be optimized to smaller. 
Moreover, we derive an analytical tradeoff relation between the preparation, measurement and the quantum coherence. We 
further experimentally test the tradeoff relation in a linear optical setup. The experimental and theoretical results are 
in good agreement and show that the quantum coherence plays a key role in bounding the resource costs in the postselected 
metrology process.
\end{abstract}

\maketitle

Parameter estimation is a fundamental subject in information theory and mathematical statistics. In quantum metrology, the 
precision can be improved beyond classical bounds by quantum effect, and the lower-bounds of variance is determined by the 
quantum Cram\'{e}r-Rao inequality~\cite{CRB1,CRB2,CRB3,CRB4},
\begin{eqnarray}
\langle(g-g_{\text{est}})^{2}\rangle\geq\frac{1}{\nu F},
\end{eqnarray}
where $g$ is the parameter to be estimated, $g_{\text{est}}$ is an unbiased estimator for $g$, $\nu$ is the number of 
independent samples, and $F$ is the quantum Fisher information (QFI) for a single sample. The quantum Cram\'{e}r-Rao 
bound describes the theoretical precision limit in parameter estimation. However, a realistic parameter estimation 
process must take into account many factors, such as the limitation in preparation of samples, the technical noise of 
detectors~\cite{T-noisy}, and the impact of decoherence~\cite{De-QIF}. Moreover, multiparameter estimation has to be 
considered in some cases~\cite{CRB5,CRB6,CRB7}.

Weak-value amplification (WVA) is a metrological protocol to amplify ultrasmall physical effects by postselection, 
and has some obvious advantages over conventional measurement under the influence of environmental noises~\cite{W1,W2,W3,W4,W5}. 
It has been widely used for precision measurements of weak signals, such as the spin Hall effect of light~\cite{WM1}, phase 
shifts~\cite{WM2,WM3}, frequency shifts~\cite{WM4}, tiny deflections of light~\cite{WM5}, and velocity measurements~\cite{WM6}. 
Under the Fisher information metric~\cite{WM7,WMX,WM8,WM9,WM10}, WVA can put all of the Fisher information of initial 
samples into a small number of post selected samples. It implies that WVA can overcome the technical noise caused by the low 
saturation of detectors. This advantage has been experimentally demonstrated in optical systems~\cite{WMa}. In Ref.~\cite{WM8}, 
the authors prove that the advantage of postselection can only be realized in nonclassical metrology.

In parameter estimation experiments, the preparation and measurement of quantum states require a certain cost. For example, the 
preparation costs can be the time or energy needed to prepare the samples, and the measurement costs can be the time or energy 
needed to reset the detector after a detection. It is important to investigate how to reduce these costs and thus obtain the 
higher accuracy of parameter estimation.

In this work, we focus on the problem about the preparation costs and measurement costs in the parameter estimation by WVA. 
The effect of WVA is determined by the pre-and post-selection states. For a general initial state, it is found that the two 
costs cannot always reach the minimum jointly in the WVA scenario. We derive a trade-off relation between the two costs and 
reveal that the two costs are constrained by the quantum coherence in the initial system state. Concretely, if and only if the 
initial state of system is prepared in the maximum coherent states, the two costs can be taken to the minimum value jointly, 
i.e., WVA performs the best advantage. While for the incoherent states, the two costs are impossibly smaller than the case in 
conventional measurement, which implies WVA has no any advantages. Consequently, quantum coherence is an indispensable resource 
to realize the technical advantage of WVA. More importantly, we give an analytical tradeoff relation between the two costs and 
the quantum coherence. Our theory is expected to have useful applications in the parameter-estimation experiments, where the 
consumption during the preparation and measurement process should be treated differently. Finally, we experimentally test the 
effect of quantum coherence on the tradeoff of the two costs in linear optical systems.

We consider that the estimation of the parameter $g$ is introduced by the unitary evolution
\begin{eqnarray}
U(g)=\exp(-igA\otimes M),
\label{q1}
\end{eqnarray}
where $M$ and $A$ are the operators of meter (denoted by $m$) and system (denoted by $s$) respectively, 
$|\Psi_{i}\rangle=|\psi_{si}\rangle|\phi_{mi}\rangle$ is the separable initial state, and $g$ is the coupling strength and 
also the parameter to be estimated.  The QFI of $g$ is defined by
\begin{eqnarray}
F=4([\partial_{g}\langle\Psi(g)|][\partial_{g}|\Psi(g)\rangle]-|[\partial_{g}\langle\Psi(g)|]|\Psi(g)\rangle|^{2})
\end{eqnarray}
where $|\Psi(g)\rangle=U(g)|\Psi_{i}\rangle$. According to the above expression, we can easily get the QFI of $g$
\begin{eqnarray}
F=4(\langle A^{2}\rangle\langle M^{2}\rangle-\langle A\rangle^{2}\langle M\rangle^{2}).
\label{q0}
\end{eqnarray}
We assume that the initial state of meter is at the balance zero point, i.e., $\langle M\rangle=0$, and the average 
of $M^{2}$ in the initial state is defined by $\Omega\equiv\langle M^{2}\rangle$. Under these conditions, we have 
$F=4\langle A^{2}\rangle\Omega$.

\emph{Parameter estimation under conventional measurement.} To begin with, we identify $[NF]^{-1}$ as the metric of the 
estimation accuracy, where $N$ is the number of the samples and here it is assumed to be large enough. Under conventional 
measurement, $R_{\text{p}}N$ is defined as the minimum preparation costs to finish the parameter estimation task, where 
$R_{\text{p}}$ is the cost to prepare one sample. $R_{\text{m}}N$ is the corresponding minimum measurement costs to detect 
all samples, where $R_{\text{m}}$ is the cost to detect one sample.

 \begin{figure}[htbp]
\centering\includegraphics[width=8cm]{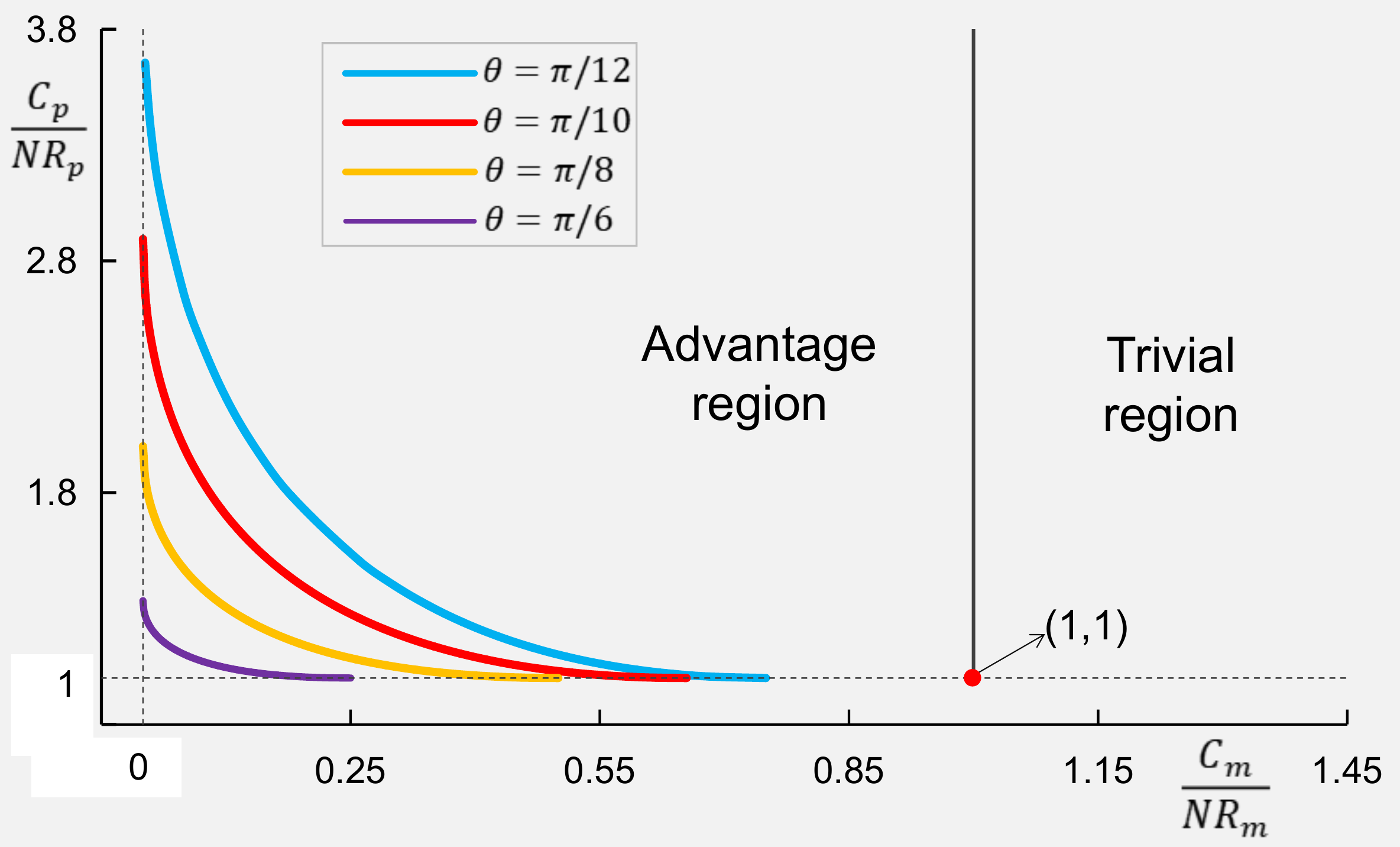}
\caption{Preparation costs and measurement costs for the initial state 
$|\psi_{i}\rangle_{s}=\cos(\theta)|\tilde{0}\rangle+\sin(\theta)|\tilde{1}\rangle$. 
The regions below the curves are forbidden.}
\label{fig1}
\end{figure}

\emph{Parameter estimation under WVA protocol.}  
Consider the unitary evolution in Eq.~(\ref{q1}), the probability to detect the system in the post-selection state 
$|\psi_{sf}\rangle$ is $p=||\langle\psi_{sf}|U(g)|\psi_{si}\rangle|\phi_{mi}\rangle||^{2}$, and the corresponding collapsed 
state of meter is
\begin{eqnarray}
|\phi_{mf}(g)\rangle=\langle\psi_{sf}|U(g)|\psi_{si}\rangle|\phi_{mi}\rangle/\sqrt{p}.
\end{eqnarray}
When $g$ is sufficiently small, the QFI of $g$ in the collapsed state $|\phi_{mf}(g)\rangle$ is
\begin{eqnarray}
F_{m}(g)=4\Omega|\text{A}_{w}|^{2}+\mathcal{O}_{F}(g),
\label{G1}
\end{eqnarray}
where $\text{A}_{w}=\langle\psi_{sf}|A|\psi_{si}\rangle/\langle\psi_{sf}|\psi_{si}\rangle$ is the weak value~\cite{WMX},
$\mathcal{O}_{F}(g)$ is the high-order terms of $g$ in the expression of the QFI $F_{m}$.
The
the postselection success probability is $p=|\langle\psi_{sf}|\psi_{si}\rangle|^{2}+\mathcal{O}_{p}(g)$, where $\mathcal{O}_{p}(g)$ 
is the high-order terms of $g$ in the expression of the probability $p$. The probabilistic QFI is
\begin{eqnarray}
f_{m}=p\cdot F_{m}=4\Delta|\langle\psi_{sf}|A|\psi_{si}\rangle|^{2}+\mathcal{O}(g).
\label{qx}
\end{eqnarray}
where $\mathcal{O}(g)$ denotes the high-order terms of $g$ which is with respect to $\mathcal{O}_{F}(g)$ and $\mathcal{O}_{p}(g)$.
In WVA, preparation costs and measurement costs are denoted by $\mathcal{C}_{p}$ and $\mathcal{C}_{m}$ respectively, and the number 
of samples needed in the WVA is denoted by $n$. To achieve the accuracy target $(NF)^{-1}$, $n$ should be satisfied $nf_{m}=NF$, 
then we can get $\mathcal{C}_{p}=nR_{p}=\frac{F}{f_{m}}R_{p}N$. The number of postselection samples is $np$ and the corresponding 
measurement costs is $\mathcal{C}_{m}=npR_{m}$. At last, we have
\begin{eqnarray}
\mathcal{C}_{p}=\frac{F}{f_{m}}R_{p}N,~~~\mathcal{C}_{m}=\frac{F}{F_{m}}R_{m}N.
\label{qa}
\end{eqnarray}

We first analyze the effect of the postselection state $|\psi_{sf}\rangle$ on $\mathcal{C}_{p}$ and $\mathcal{C}_{m}$.
The optimal state $|\psi_{sf}^{\text{opt}}\rangle$, which is defined as $|\psi_{sf}^{\text{opt}}\rangle=A|\psi_{si}\rangle
/\sqrt{\langle A^{2}\rangle}$~\cite{WMX}, can make $f_{m}(|\psi_{sf}^{\bot}\rangle)$ take the maximum, which can approximately 
equal to $F$, but cannot be larger than it \cite{WMX}. Then, the minimum value of $\mathcal{C}_{p}$ satisfies 
$\mathcal{C}_{p}(|\psi_{sf}^{\text{opt}}\rangle)\gtrapprox R_{p}N$, but the measurement costs 
$\mathcal{C}_{m}(|\psi_{sf}^{\text{opt}}\rangle)$ may not achieve the minimum.
The near-orthogonal state $|\psi_{sf}^{\bot}\rangle$, which is defined as $\langle\psi_{si}|\psi_{sf}^{\bot}\rangle\approx0$, can 
make $\mathcal{C}_{m}(|\psi_{sf}^{\bot}\rangle)$ take the minimum value since the weak value $A_w$ in Eq.~(\ref{G1}) can approach 
its maximum at the near-orthogonal state $|\psi_{sf}^{\bot}\rangle$. However, the preparation costs 
$\mathcal{C}_{p}(|\psi_{sf}^{\bot}\rangle)$  cannot reach its minimum.

\begin{figure}[tp]
\centering
\includegraphics[width=8.8cm]{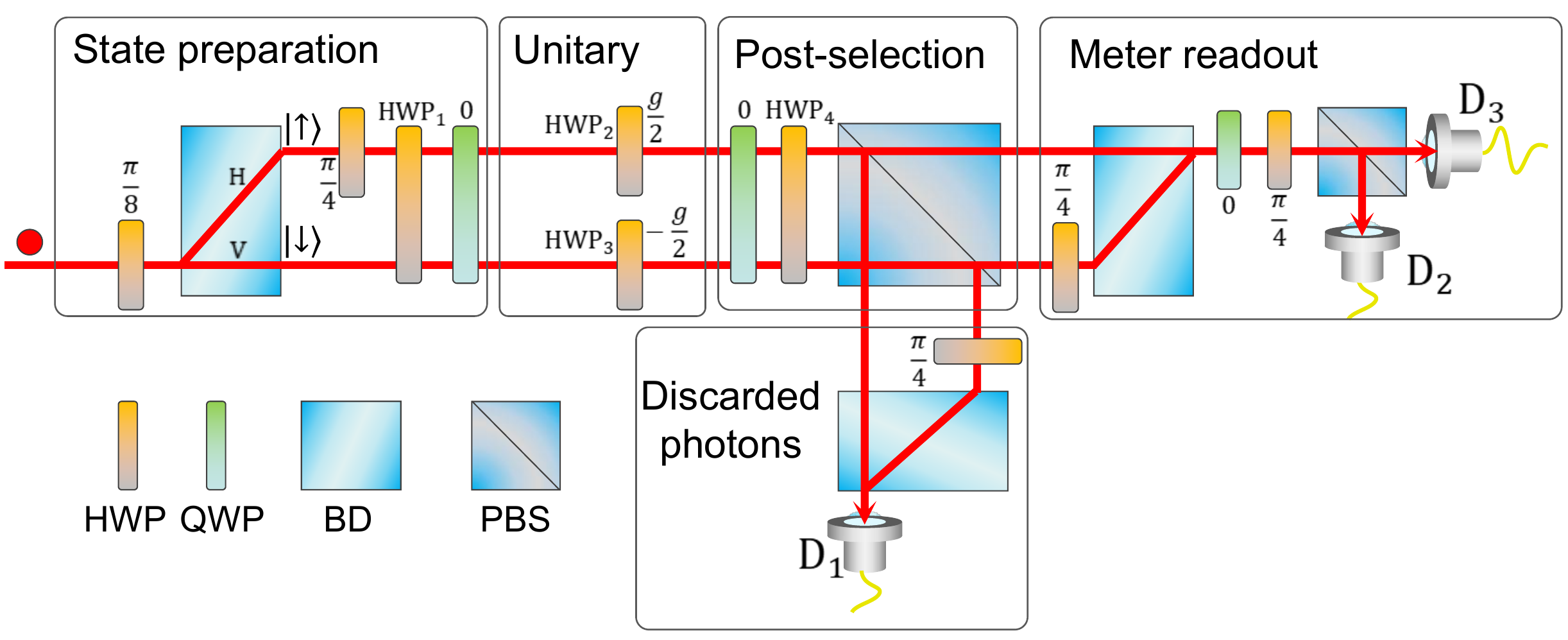}
\caption{Experimental implementation to estimate a polarization change via WVA. 
The angles of all QWPs are set to 0. The devices are half-wave plates (HWP), 
quarter-wave plates (QWP), beam displacers (BD) and polarizing beam splitters (PBS).
}
\label{fig2}
\end{figure}

\begin{figure*}[tp]
\centering
\includegraphics[width=17cm]{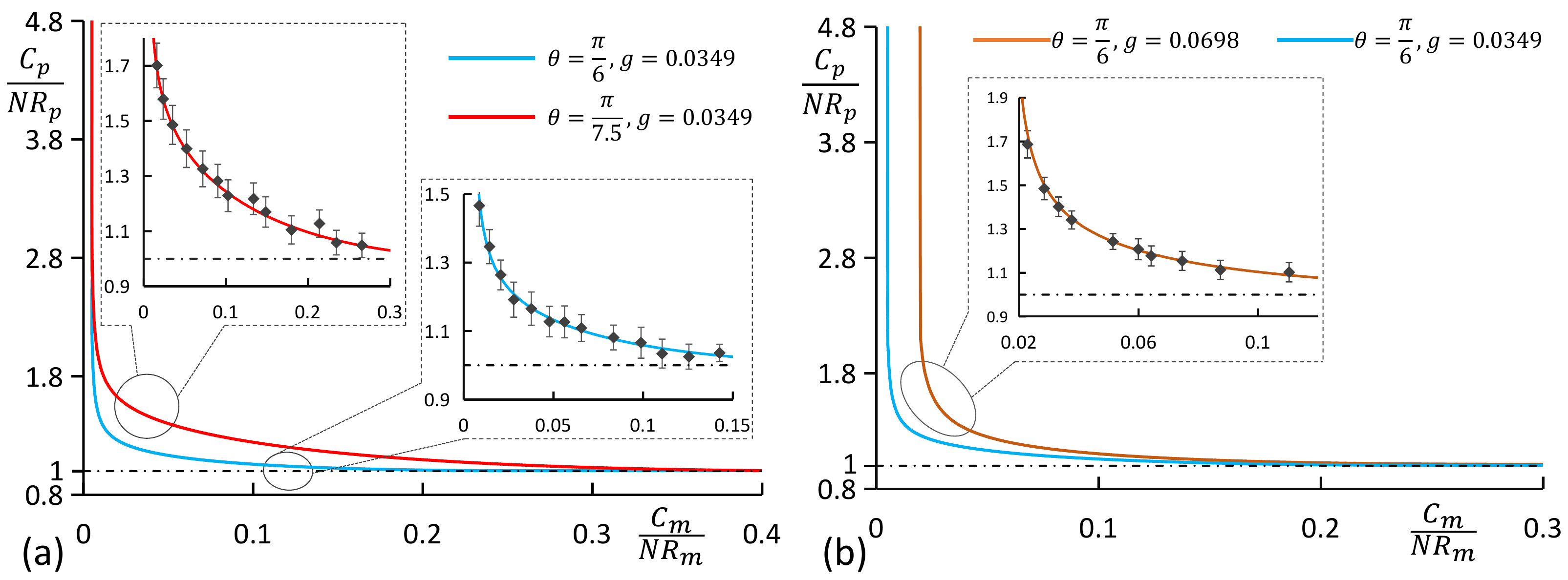}
\caption{Experimental data of the tradeoff bound of $\mathcal{C}_{p}$ and $\mathcal{C}_{m}$ 
for the initial state $|\psi_{si}\rangle=\cos(\theta)|\tilde{0}\rangle+\sin(\theta)|\tilde{1}\rangle$.
The experimental data are denoted by rhombuses. Solid lines correspond to theoretical results. 
(a) Different parameters to be estimated, i.e., $g=0.0349~\text{rad}$ and $0.0698~\text{rad}$. The 
parameter $\theta=\pi/6$. In the experiment, $\sigma=\sigma_{y}$ with the eigenstates 
$|\tilde{0}\rangle=1/\sqrt{2}(|H\rangle+i|V\rangle)$ and $|\tilde{1}\rangle=1/\sqrt{2}(|H\rangle-i|V\rangle)$, 
$M=\sigma_{z}$ with the eigenstates $|\uparrow\rangle$ and $|\downarrow\rangle$.}
\label{fig3}
\end{figure*}
To minimize $\mathcal{C}_{p}$ and $\mathcal{C}_{m}$ jointly, the initial state of system should be prepared in the special states. 
We consider that the system $s$ is a qubit system and let $A=\sigma$ which is one of the Pauli operators, then 
$|\psi_{sf}^{\text{opt}}\rangle=\sigma|\psi_{si}\rangle$ and $F=4\Omega$ since $\langle\sigma^{2}\rangle=1$.
We define the eigenstates of $\sigma$ are $|\tilde{0}\rangle,|\tilde{1}\rangle$ with the eigenvalues $1$ and $-1$.
For a general two-dimensional state, the density operator can be represented by $\rho=\frac{1}{2}(I+\textbf{r}\cdot\boldsymbol{\sigma})$, 
where $I$ is the $2\times2$ identity matrix, $\textbf{r}=(r_{1},r_{2},r_{3})$ is the Bloch vector, and 
$\boldsymbol{\sigma}=(\sigma_{x},\sigma_{y},\sigma_{z})$ is the vector of Pauli matrices. For two pure states $|a\rangle$ and 
$|b\rangle$, the module square of the overlap $|\langle a|b\rangle|^2$ can be written as
\begin{eqnarray}
|\langle a|b\rangle|^2=\cos^{2}\left(\frac{1}{2}\Theta(\textbf{r}_{a},\textbf{r}_{b})\right),
\end{eqnarray}
where $\Theta(\textbf{r}_{a},\textbf{r}_{b})=\arccos(\textbf{r}_{a}\cdot\textbf{r}_{b})$ is the angle between $\textbf{r}_{a}$ 
and $\textbf{r}_{b}$. When the high-order of $g$ in Eq.~(\ref{qa}) can be neglected, $\mathcal{C}_{p}$ and $\mathcal{C}_{m}$ can 
be written as
\begin{eqnarray}
\begin{split}
&\mathcal{C}_{p}=\frac{1}{\cos^2(\Theta(\textbf{r}_{1},\textbf{r}_{2})/2)}R_{p}N,&\\
&\mathcal{C}_{m}=\frac{\mathcal{C}_{p}}{R_{p}}\cos^2(\Theta(\textbf{r}_{1},\textbf{r}_{3})/2)R_{m},&
\end{split}
\end{eqnarray}
where $\textbf{r}_{1}$,$\textbf{r}_{2}$,$\textbf{r}_{3}$ are Bloch vectors of 
$|\psi_{sf}\rangle$,$\sigma|\psi_{si}\rangle$,$|\psi_{si}\rangle$.
The tradeoff relation of $\mathcal{C}_{p}$ and $\mathcal{C}_{m}$ can be obtained based on the
geometric relationship $|\Theta(\textbf{r}_{1},\textbf{r}_{2})-\Theta(\textbf{r}_{1},\textbf{r}_{3})|\leq 
\Theta(\textbf{r}_{2},\textbf{r}_{3})$, then we have
\begin{eqnarray}
\begin{split}
&\Big|2\arccos(\sqrt{\frac{R_{p}N}{\mathcal{C}_{p}}})-2\arccos(\sqrt{\frac{\mathcal{C}_{m}R_{p}}{\mathcal{C}_{p}R_{m}}})\Big|&\\
&\leq2\arccos(\sqrt{1-\text{C}_{l_{1}}(|\psi_{si}\rangle)}),&
\end{split}
\label{q3}
\end{eqnarray}
where $\text{C}_{l_{1}}(|\psi_{si}\rangle)$ is the $l_{1}$-norm of coherence of the initial state under the reference basis 
$\{|\tilde{0}\rangle,|\tilde{1}\rangle\}$. For a quantum state $\rho$, $\text{C}_{l_{1}}$ can be calculated by 
$\text{C}_{l_{1}}(\rho)=\sum_{i\neq j}|\rho_{ij}|$, where $\rho_{ij}$ is the element of the density matrix in the space spanned 
by the reference bases $\{|\tilde{0}\rangle,|\tilde{1}\rangle\}$.

From Eq.~(\ref{q3}), we find that the minimum value of $\mathcal{C}_{m}$ is restricted as $\mathcal{C}_{m}=1-\text{C}_{l_{1}}^{2}$
when $\mathcal{C}_{p}$ takes the minimum value $\mathcal{C}_{p}=NR_{p}$.
Obviously, if and only if the initial state is the maximally coherent state, i.e.,
$\text{C}_{l_{1}}=1$, the minimum values of $\mathcal{C}_{p}$ and $\mathcal{C}_{m}$ can be reached simultaneously.
As shown in Fig.\ref{fig1}, we demonstrate the bound on $\mathcal{C}_{p}$ and $\mathcal{C}_{m}$ for the initial states with 
different amounts of coherence resource. It should be noted that the condition of WAV should always be satisfied in Fig.~\ref{fig1}, 
i.e., $g|\text{A}_{w}|\Omega\ll1$, and the measurement costs $\mathcal{C}_{m}$ can not be zero in experiments. On the other hand, 
for any incoherent states, i.e., $\rho_{si}=\mu|\tilde{0}\rangle\langle\tilde{0}|+(1-\mu)|\tilde{1}\rangle\langle\tilde{1}|$, we 
have $F=4\Omega$ (see details in the Appendix). It can be proved that the QFI in the post-selection state cannot exceed F when the 
initial system state is incoherent, i.e., $F_{m}\leq F$ (the details are shown in the Appendix). It implies the measurement cost 
$\mathcal{C}_{m}$ is larger than that in the conventional scenario, i.e., $\mathcal{C}_{m}\geq R_{m}N$. In this case, the WAV 
scenario cannot provide any advantages. Hence, we define the left side of $\frac{\mathcal{C}_{m}}{NR_{m}}=1$ as the advantage 
region while the right side as the trivial region.

\emph{Experimental setup.} 
To demonstrate the effect of quantum coherence in the WVA scenario of parameter estimation, we experimentally estimate a tiny 
polarization angle change of a single photon. The experimental setup is sketched in Fig.~\ref{fig2}. A single-photon source is 
produced via a spontaneous parametric down-conversion (SPDC) process by pumping a 1.5-cm-long type-II periodically poled 
potassium titanyl phosphate (PPKTP) nonlinear crystal with ultraviolet pulses at a 405\,nm centered wavelength. One photon is 
directly detected as a trigger. The other one is prepared in a pure state of the spatial modes and the polarization modes.

We employ the horizontal and vertical polarization modes $|H\rangle$ and $|V\rangle$ as the system basis vectors. And the spatial 
modes $|\!\uparrow\rangle$ and $|\!\downarrow\rangle$ act as the basis vectors of the meter. In the state preparation, the state 
$|\phi_{mi}\rangle=\frac{1}{\sqrt{2}}(|\uparrow\rangle+|\downarrow\rangle)$ of the meter is prepared by adjusting the first HWP 
to $\pi/8$. The state $|\psi_{si}\rangle=\cos(\theta)|\tilde{0}\rangle+\sin(\theta)|\tilde{1}\rangle$ of polarization can be 
prepared by adjusting the $\text{HWP}_{1}$ to $\pi/8-\theta/2$, where $|\tilde{0}\rangle=1/\sqrt{2}(|H\rangle+i|V\rangle)$ and 
$|\tilde{1}\rangle=1/\sqrt{2}(|H\rangle-i|V\rangle)$. The unitary evolution is $U=\exp(ig\sigma_{y}^{s}\otimes\sigma_{z}^{m})$, 
where $\sigma_{y}^{s}=i(|H\rangle\langle V|-|V\rangle\langle H|)$, $\sigma_{z}^{m}=|\uparrow\rangle\langle\uparrow|
-|\downarrow\rangle\langle\downarrow|$, and the change of $g$ is implemented by rotating the angles of $\text{HWP}_{2}$ and 
$\text{HWP}_{3}$ to $g/2$ in opposite directions. Then, the polarization mode is projected onto $|\psi_{sf}\rangle$ by postselecting 
the photon coming from the direct path of the PBS, and the post-selection state$|\psi_{sf}\rangle=\cos(\alpha)|\tilde{0}\rangle
+\sin(\alpha)|\tilde{1}\rangle$ is determined by rotating $\text{HWP}_{4}$. To detect the probability of successful post-selection, 
the reflected discarded photons are measured by the $D_{1}$ detector. Finally, the meter readout is provided by the detectors 
$D_{2}$ and $D_{3}$.

\emph{Experimental result.} 
We demonstrate the cost ratios between conventional measurement and weak measurement with the initial system state 
$\cos(\theta)|\tilde{0}\rangle+\sin(\theta)|\tilde{1}\rangle$. According to Eq.~(\ref{qa}), we experimentally represent 
$C_{p}/[NR_{p}]$ and $C_{m}/[NR_{m}]$ by detecting $p$ and $F_{m}$. The Cramer-Rao bound is asymptotically attainable 
by the maximum likelihood estimator, thus the Fisher information is approximately obtained by measuring the variance of 
$g_{\text{est}}$, i.e., $F_{m}\simeq[\nu\delta^{2}g_{\text{est}}]^{-1}$. We detect about $700$ post-selection photons 
and collect the measurement outcomes from the two detectors $D_{2}$ and $D_{3}$. Then the estimator $g_{\text{est}}$ is 
obtained with the maximum likelihood, which maximizes the posterior probability based on the obtained data. The above 
process is repeated 1000 times to get the distribution and the variance of $g_{\text{est}}$.

In Fig.~\ref{fig3}(a), the experimental and theoretical results are in good agreement and show the tradeoff bound between 
$\mathcal{C}_{p}$ and $\mathcal{C}_{m}$ for the superposition coefficients $\theta=\pi/6$ and $\theta=\pi/7.5$ with 
$g=0.0349~\text{rad}$. The regions below the curves are forbidden. If $|\tilde{0}\rangle$ and $|\tilde{1}\rangle$ are 
defined as the reference bases, the $l_{1}$-norm of coherence is $\text{C}_{l_{1}}(|\psi_{si}\rangle)=\sin(2\theta)$, 
and the more coherence resource of the initial state $(\theta\rightarrow\pi/4)$ would give the lower cost. The curves 
in Fig.~\ref{fig3}(a) is a little different from that in Fig.\ref{fig1}, where $\mathcal{C}_{m}$ can not be infinitely 
close to 0. This is because $g$ is a finitely small parameter, and the regime of validity of the weak-value theory can 
not be satisfied when $\text{A}_{w}$ is too large. In Fig.~\ref{fig3}(b), we compare the tradeoff bound with $g=0.0349~\text{rad}$ 
and $0.0698~\text{rad}$ by setting $\theta=\pi/6$. The tradeoff bound provided by Eq.~(\ref{q3}) becomes more tight when 
$g$ is smaller.


\emph{Conclusion.} 
In the actual parameter estimation process, it is necessary to consider the consumption of preparation resources and measurement 
resources. In this work, we investigated the problem of the resource cost in the weak-value amplification. We derive a tradeoff 
relation between the preparation costs, measurement costs, and the quantum coherence of the initial system state. It reveals a 
noteworthy fact that the minimum costs of preparation and measurement are not independent in the WVA scenario. Moreover, they are 
bounded by the quantum coherence of the system. The theoretical and experimental results agree well and show that increased quantum 
coherence can reduce the preparation cost and measurement cost jointly. For the initial state with the imperfect coherence resource, 
we give the optimal strategy to complete the parameter estimation task with the different consumptions of preparation and measurement. 
We hope that this work will provide some useful assistance in the design of various sensors. Furthermore, our work can be used to explain 
the advantage of coherent sources in optical super-resolution~\cite{WMa1,WMa2,WMa3}, where the coherence can help to reduce the measurement 
costs.

\begin{acknowledgments}
This work was supported by the National Natural Science Foundation of China (11935012, 12175052, 11775065, 12205092 and 12175075).
\end{acknowledgments}

\appendix

\section{appendix}
For an initial state in the spectral decomposition form $\rho_{in}=\sum_{i=1}^{M}\lambda_{i}|\psi_{in,i}\rangle\langle\psi_{in,i}|$, 
the QIF in the final state $U(g)\rho_{in}U^{\dagger}(g)$ is
\begin{eqnarray}
\begin{split}
F(g)&=\sum_{i=1}^{M}\frac{(\partial_{g}\lambda_{i})^{2}}{\lambda_{i}}+
\sum_{i=1}^{M}4\lambda_{i}\langle\psi_{in,i}|(\partial_{g}U^{\dagger})(\partial_{g}U)|\psi_{in,i}\rangle&\\
&-\sum_{i,j=1}^{M}\frac{8\lambda_{i}\lambda_{j}}{\lambda_{i}+\lambda_{j}}\big|\langle\psi_{in,i}|U^{\dagger}
\partial_{g}U|\psi_{in,i}\rangle\big|^{2}.&
\end{split}
\end{eqnarray}
Now, we calculate $F$ and $F_{m}$ for the initial state 
$\rho_{si}\otimes|\phi_{mi}\rangle\langle\phi_{mi}|=(\mu|\tilde{0}\rangle\langle\tilde{0}|+(1-\mu)|\tilde{1}\rangle
\langle\tilde{1}|)\otimes|\phi_{mi}\rangle\langle\phi_{mi}|$. After the unitary evolution $U(g)=\exp(-ig\sigma\!\otimes\!M)$, we have
\begin{eqnarray}
\rho_{sm}(g)=U(g)\big[\rho_{si}\otimes|\phi_{mi}\rangle\langle\phi_{mi}|\big]U^{\dagger}(g).
\end{eqnarray}
Since $\langle\phi_{mi}|M|\phi_{mi}\rangle=0$, we have
\begin{eqnarray}
\langle\tilde{x}|\langle\phi_{mi}|U^{\dagger}\partial_{g}U|\tilde{y}\rangle|\phi_{mi}\rangle=i\langle\tilde{x}|\sigma|\tilde{y}\rangle
\langle\phi_{mi}|M|\phi_{mi}\rangle=0.
\end{eqnarray}
where $\tilde{x}=\tilde{0},\tilde{1};~\tilde{y}=\tilde{0},\tilde{1}$. At last, we can get
\begin{eqnarray}
\begin{split}
F(\rho_{sm}(g))&=4(\mu\langle\tilde{0}|\sigma^{2}|\tilde{0}\rangle+(1-\mu)\langle\tilde{1}|\sigma^{2}|\tilde{1}\rangle)
\langle\phi_{mi}|M^{2}|\phi_{mi}\rangle&\\
&=4\Omega.&
\end{split}
\end{eqnarray}

With an auxiliary system $a$, the initial state can be written as $\big(\sqrt{\mu}|\tilde{0}\rangle|0_{a}\rangle
+\sqrt{1-\mu}|\tilde{1}\rangle|1_{a}\rangle\big)|\phi_{mi}\rangle$. After the unitary evolution and postselection, the 
collapsed state of the meter and the auxiliary system is
\begin{eqnarray}
|\Psi_{am}(g)\rangle=\xi|0_{a}\rangle(e^{igM}|\phi_{mi}\rangle)+\eta|1_{a}\rangle(e^{-igM}|\phi_{mi}\rangle)
\end{eqnarray}
where $\xi=\frac{\langle\psi_{sf}|\tilde{0}\rangle}{\sqrt{u|\langle\psi_{sf}|\tilde{0}\rangle|^{2}}}$ and 
$\eta=\frac{\langle\psi_{sf}|\tilde{1}\rangle}{\sqrt{(1-\mu)|\langle\psi_{sf}|\tilde{1}\rangle|^{2}}}$. We can easily get 
$F(|\Psi_{am}(g)\rangle)=4\Omega$. Thus, after tracing the auxiliary system, the QIF in the meter can not exceed $4\Omega$.

\end{document}